\documentclass[epsf,twocolumn,showpacs,preprintnumbers]{revtex4}

\usepackage{amsmath,amssymb,amsfonts}
\usepackage{graphics}
\usepackage{dcolumn} 
\usepackage{bm,graphicx,color}
\usepackage{epsfig,comment}
\begin{document}
\title{The dawn of the nickel age of superconductivity}
\author{Warren E. Pickett$^*$}
\affiliation{University of California Davis, Davis CA 95616} 
\email{pickett@physics.ucdavis.edu}
\date{\today}
\begin{abstract}
A Commentary in: Nature Reviews Physics {\bf 3}, 7 (2021). doi.org/10.1038/s42254-020-00257-3

\vskip 7mm
\centering{{\bf Key Advances}}
\begin{itemize}
\item Extraction of layers of oxygen ions from thin films of perovskite structure 
(Nd,Sr)NiO$_2$ has led to emergence of the long sought nickelate superconductivity.
\item Comparison of the nickelate with the isostructural and isovalent, high temperature superconducting, cuprate has framed the main debate on the importance of their obvious similarities and less evident differences.
\item Several observations —- electronic, magnetic, and structural —- point toward   a new frontier Ni $d_{z^2}$  orbital, with its role in interlayer coupling, 
in addition to the $d_{x^2-y^2}$  orbital that has dominated the discussion in cuprate superconductivity.
\end{itemize} 
\end{abstract}
\maketitle


Whereas high-temperature superconductivity in cuprates has been studied for 30 years, during the past year it has been reported for the first time in nickelates. This raises new questions for materials physicists and chemists about the mechanism of superconductivity — despite the electronic similarities of Cu and Ni, it seems that nickelate superconductivity requires consideration of  a second orbital.

Not long after the discovery of high-temperature superconductivity in cuprates in the early 1990s, it was suggested that isostructural nickelates might have similar properties. After all, Ni borders Cu in the periodic table, with its Ni$^{1+}$ ion 
being isoelectronic ($d^9$) with the Cu$^{2+}$ ion. 
Three decades lapsed with substantial study of nickelates, but no fruit borne. 
The drought ended last year with the announcement of T$_c$ up to 15K in a 
nickelate, NdNiO$_2$\cite{DLi2019} and the confirmation of the 
discovery\cite{Zeng2020} earlier 
this year. But old and now new questions remain to be answered.

The discovery of superconductivity in a nickelate compound is anticipated to clarify the underlying mechanism(s) of superconductivity in their sister cuprate materials. One of the most prominent questions is whether they are two sides of the same coin, given the same structure and similar valence, or whether they involve fundamentally different electronic behaviour.

\begin{figure*}[!htbp]
\centerline{\includegraphics[width=1.6\columnwidth]{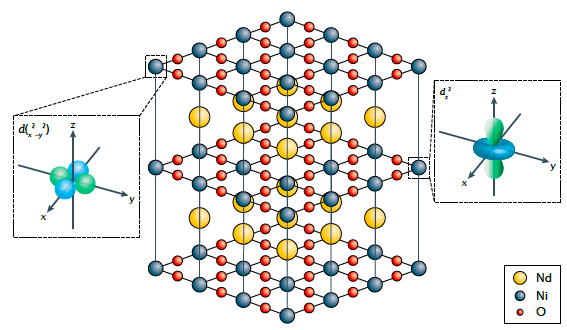}}
\caption{ Infinite layer nickelate, NdNiO$_2$.  A schematic showing
the structure of an infinite layer nickelate, illustrating also the two
active Ni orbitals. Red spheres indicate oxygen, gold spheres are Nd.
The sample used in Ref. 1 was synthesized in thin film form by
chemically extracting layers of oxygen atoms from a layer-by-layer
synthesized thin perovskite film. The most evident effect of the
synthesis technique was that it stabilized an `infinite layer' structure
in which the layer of oxygen atoms linking CuO$_2$ layers in most
cuprates is missing. Hole-doped (Nd,Sr)NiO$_2$ seems to mimic
isostructural and isovalent (Ca,Sr)$_{1-x}$CuO$_2$, which
superconducts when hole-doped up to 110K.\cite{Azuma1992}}
\label{figure1}
\end{figure*}

Besides the factor-of-seven larger T$_c$ achieved so far in the 
cuprate,\cite{Azuma1992} 
there are several differences between the compounds, including different 
lattice constants (atomic sizes). Noted by several theory groups is that 
Nd (or La, which is an often used substitute because it has no open $4f$ 
shell) has a $5d$ band that contributes electronic carriers. Another 
difference is that the interlayer Nd$^{3+}$ ion provides a different 
Madelung potential and crystal field on the metal ion compared to the 
Ca$^{2+}$ ion. The primary effects are that the metal $d_{z^2}$ orbital 
becomes active in NdNiO$_2$ whereas it is a bystander in the cuprate, 
and the $2p$ states in O are more tightly bound by 1 eV or more in the 
nickelate relative to the cuprate.\cite{Botana2020} 

This dichotomy of viewpoint had appeared already two decades earlier in 
comparisons of LaNiO$_2$ and CaCuO$_2$.  A 1999 study\cite{Anisimov1999} 
emphasized the 
similarities. Besides structure and metal ion valence $d^9$ configuration, 
the position and bandwidth of the crucial $d_{x^2-y^2}$  band is similar. 
Yet there are qualitative differences in the properties of the undoped 
materials. CaCuO$_2$ is an insulating antiferromagnet whereas NdNiO$_2$ 
displays no magnetic order and is conducting —- albeit weakly so —- 
like so many other oxide `bad metals', several of which superconduct.  
A 2004 study\cite{KWLee2004} pointed out two further substantial differences. The 
tripositive rare earth $5d$ orbitals in LaNiO$_2$ become partially 
occupied, thereby hole-doping the Ni $3d$ bands already at stoichiometry, 
and providing screening of Ni moments as well as competing itinerant 
magnetic exchange processes.  An additional difference is the substantial 
role played by the Ni $d_{z^2}$ orbital in the Fermi level region.

This unusual involvement of the Ni $d_{z^2}$ orbital introduces new 
physics. This interference can be traced to two factors. First, in the 
infinite-layer structure (Fig. 1), the missing oxygen layer results in 
a 15\% decrease of the $\hat c$ lattice parameter relative to cubic 
perovskite. As a result, the $d_{z^2}$  orbital is less isolated 
concerning interlayer coupling, but this aspect should be similar in the 
cuprate and nickelate. In addition, the rare earth (Nd, or La) $5d$ 
orbital is larger than the $3d$ orbital of Ca. The result is enhanced 
$\hat c$ axis coupling, but also interference by the rare earth ion 
in the behaviour of the $e_g$ orbitals of the metal ion. The 
interference is via crystal field, ${\hat c}$-axis coupling, and by 
providing $5d$ carriers that interact with the Ni moments.

In both materials the $d_{z^2}$  orbital is essentially degenerate 
with the $t_{2g}$ orbitals, but in the nickelate its separation from 
the $d_{x^2-y^2}$  energy is only 70\% as large\cite{Botana2020}. These numbers are 
obtained from density functional calculations used for weakly correlated 
materials, which serve as the accepted `underlying electronic structure' 
of quantum materials. A point of wide agreement is that intra-atomic 
Ni interaction effects (as in the cuprate) will be important in the 
calculations, these being the Coulomb repulsion, $U$, between electrons 
and Hund's coupling, $J$, between parallel spin electrons, in the $3d$ 
shell. Values of $U$ in the 3-9 eV range are being used in various 
theoretical treatments, and the actual value seems to be a more important 
issue in the nickelate than in the cuprate. An aspect of uncertainty 
is the degree of screening by the charged carriers and its effect on 
the two $e_g$ orbitals.

An open question concerns the status of the configuration of the 
Ni$^{1+}$ ion. In the cuprates, the $d^9$ state in the Cu$^{1+}$ ion, 
with spin $S$=1/2 in the $d_{x^2-y^2}$  orbital, is the unambiguous 
reference state. In doped cuprate samples, the $d^8$ state becomes 
relevant, and hybridization with $p_{\sigma}$ orbitals on the four 
neighboring O ions must be considered. Although Ni$^{1+}$ ion appears 
analogous to the Cu$^{1+}$ ion, hybridization is reduced and the 
supporting players upon doping have been suggested to be different. 

One suggestion, based on an interacting cluster model, is that the 
$d^8$ configuration, such as Ni$^{2+}$, comes more strongly into 
play.\cite{Jiang2020} This ion can assume either the prosaic and symmetric $S$=0 
state, or the conventional Hund's rule $S$=1 state, which has been 
proposed as nearby in energy. The non-magnetic $S$=0 configuration 
has the $pd\sigma$-bonding $d_{x^2-y^2}$  orbital of both spins 
occupied (or unoccupied, depending on viewpoint), and minimizes 
spin physics to the extent it is involved. $S$=1 contributions would 
enrich the `spin physics'. These issues impact the follow-up 
dynamical theory, which with all relevant degrees of freedom 
(yet to be agreed on) should contain the real physics of nickelates.

A further possibility in this local-ion ansatz is the distinctive 
inter-orbital $S^*$=0 `off-diagonal singlet' with electrons (or holes) 
of opposite spins in each of the $e_g$ orbitals —- the 
$d_{x^2-y^2}^{\uparrow}$  $d_{z^2}^{\downarrow}$ configuration -— 
giving a spin singlet with internal orbital structure and anisotropic 
interorbital coupling. This possibility seems radical as it violates 
Hund's first rule. The $S^*$ singlet was found already in one of the 
early studies\cite{KWLee2004} but at an interaction strength $U$ thought to be 
unphysically large. Unlike the cuprate, even at unrealistically 
large values of $U$, the nickelate never becomes insulating 
(pure $d^9$).\cite{KWLee2004} This off-diagonal singlet does at least 
neighbour the physical range of the doped nickelate, in which case 
it would enter theories treating the dynamics.

Given the unsettled nature of the electronic structure of (Nd,Sr)NiO$_2$, 
it is pertinent to note the recently synthesized\cite{Matsumoto2019} (under pressure) 
$d^8$ compound Ba$_2$NiO$_2$(AgSe)$_2$ (BNOAS). Its structure has a 
NiO$_2$ `infinite layer' interspersed with an insulating AgSe spacer 
layer, separated by a Ba$^{2+}$ layer. Other compounds with a similar
 Ni$^{2+}$O$_2$ layer are nonmagnetic.\cite{Matsumoto2019} BNOAS shows no 
Curie–Weiss moment in susceptibility, yet displays a magnetic transition 
(also in susceptibility) around 130 K. Calculations using two different 
codes\cite{Jin2020} obtain a small gap insulating state for BNOAS, with the same $S^*$=0 
singlet as the ground state of the $d^8$ ion — spin-polarized internally 
but with no net moment. This unanticipated state is further evidence 
that the off-diagonal singlet discussed above is `near' in energy to 
the Ni ground state in doped NdNiO$_2$.

Dynamical treatment of the Ni intra-atomic repulsion $U$ is required to 
address spectroscopic data more directly. Several such studies from 
single band to six band models, which differ in methods and in choices 
of $U$, do not arrive at a consensus on several spectroscopic features. 
They do generally concur that the $d^9$ configuration is dominant, 
somewhat at odds with the cluster calculation\cite{Jiang2020} 
and correlated band 
result\cite{Jin2020} that suggest the importance of $d^8$ 
($S$=0 or $S^*$=0 singlets). 
At stoichiometry, the correlated Fermi surface remains the same as in 
density functional theory calculations and the low energy band structure 
is renormalized. At physical doping levels and for binding energy larger 
than 1 eV the excitations become incoherent (ill-defined). 

A crucial question is why bulk-prepared Nd$_{0.8}$Sr$_{0.2}$NiO$_2$ is 
not superconducting\cite{Wang2020} whereas the thin films are. 
The most evident difference, a 3\% larger $c$-axis lattice constant, 
provides additional evidence of the involvement and sensitivity of the 
Ni $d_{z^2}$  orbital, which provides a new frontier orbital that is 
active for nickelate superconductivity.

\section{Acknowledgments}
The author acknowledges collaboration with K.-W. Lee, and several exchanges with A. S. Botana. This work was supported by National Science Foundation Grant DMR 1207622. For the computations we used the Extreme Science and Engineering Discovery Environment (XSEDE), which is supported by National Science Foundation grant number ACI-1548562.


\end{document}